\documentclass[sigconf, nonacm]{acmart}
\usepackage{algorithm}
\usepackage{algpseudocode}
\usepackage{natbib}
\usepackage{multirow}
\usepackage{geometry}
\usepackage{graphicx} 
\usepackage{booktabs} 
\usepackage{lscape} 
\usepackage{todonotes}
\usepackage{threeparttable}
\newcommand\vldbdoi{XX.XX/XXX.XX}
\newcommand\vldbpages{XXX-XXX}
\newcommand\vldbvolume{14}
\newcommand\vldbissue{1}
\newcommand\vldbyear{2020}
\newtheorem{problem}{Problem}
\newtheorem{definition}{Definition}%
\newcommand\vldbauthors{\authors}
\newcommand\vldbtitle{\shorttitle} 
\newcommand\vldbavailabilityurl{URL_TO_YOUR_ARTIFACTS}
\newcommand\vldbpagestyle{plain} 

\begin{document}
\title{Towards Practical Benchmarking of Data Cleaning Techniques: On Generating Authentic Errors via Large Language Models}


\author{Xinyuan Liu$^{1}$, Jiahui Chen$^{1}$, Bocheng Hu$^{1}$, Yu Sun$^{1}$, \\ Xinyang Chen$^{2}$, Shaoxu Song$^{3}$, Yongxin Tong$^{4}$}
\affiliation{%
  \institution{\hspace{0em} $^1$Nankai University \hspace{1em}  $^2$Harbin Institute of Technology, Shenzhen \hspace{1em} $^3$Tsinghua University \hspace{1em} $^4$Beihang University \\
  \{2112614@mail., 2110694@mail., 2111194@mail., sunyu@\}nankai.edu.cn \quad chenxinyang@hit.edu.cn \quad sxsong@tsinghua.edu.cn \quad  yxtong@buaa.edu.cn}
}







\begin{abstract}
Data quality remains an important challenge in data-driven systems, as errors in tabular data can severely compromise downstream analytics and machine learning performance. Although numerous error detection algorithms have been proposed, the lack of diverse, real-world error datasets limits comprehensive evaluation. Manual error annotation is both time-consuming and inconsistent, motivating the exploration of synthetic error generation as an alternative. In this work, we introduce TableEG, a framework that leverages large language models (LLMs) to generate authentic errors. By employing a table fine-tuning strategy and a triplet representation \((I, T, O)\) to model error generation, detection, and correction tasks, TableEG captures the complex dependencies inherent in two-dimensional tables. Trained on 12 real-world datasets spanning 10 diverse domains, TableEG ensures that the synthesized errors faithfully reflect authentic error distributions. Experimental results indicate that errors generated by TableEG exhibit superior pattern and distribution similarity compared to both rule-based methods and LLM-generated errors without fine-tuning. Furthermore, performance metrics on TableEG-generated errors closely align with those on real-world errors across nearly all datasets and detection algorithms, particularly for machine learning based detection techniques. Overall, TableEG not only bridges the gap between synthetic and real-world errors but also establishes a robust benchmark for subsequent error detection and correction tasks.
\end{abstract}

\maketitle

\pagestyle{\vldbpagestyle}
\begingroup\small\noindent\raggedright\textbf{PVLDB Reference Format:}\\
\vldbauthors. \vldbtitle. PVLDB, \vldbvolume(\vldbissue): \vldbpages, \vldbyear.\\
\href{https://doi.org/\vldbdoi}{doi:\vldbdoi}
\endgroup
\begingroup
\renewcommand\thefootnote{}\footnote{\noindent
This work is licensed under the Creative Commons BY-NC-ND 4.0 International License. Visit \url{https://creativecommons.org/licenses/by-nc-nd/4.0/} to view a copy of this license. For any use beyond those covered by this license, obtain permission by emailing \href{mailto:info@vldb.org}{info@vldb.org}. Copyright is held by the owner/author(s). Publication rights licensed to the VLDB Endowment. \\
\raggedright Proceedings of the VLDB Endowment, Vol. \vldbvolume, No. \vldbissue\ %
ISSN 2150-8097. \\
\href{https://doi.org/\vldbdoi}{doi:\vldbdoi} \\
}\addtocounter{footnote}{-1}\endgroup

\ifdefempty{\vldbavailabilityurl}{}{
\vspace{.3cm}
\begingroup\small\noindent\raggedright\textbf{PVLDB Artifact Availability:}\\
The source code, data, and/or other artifacts have been made available at \url{https://github.com/viviancircle/TableEG}.
\endgroup
}

\section{Introduction}
\label{sec:introduction}


Ensuring high data quality is a fundamental challenge in data-driven systems, as erroneous tabular data can terribly degrade the reliability of analytical processes and machine learning models~\cite{rahm2000data, chu2016data}. To tackle these errors, researchers spend a lot of time designing various data cleaning techniques \cite{chu2013holistic,mahdavi2019raha,rezig2021horizon}. As we know, sufficient training data with labeled errors are important for the performance of data cleaning models \cite{rekatsinas2017holoclean}. Unfortunately, the manual labeling is very expensive, and obtaining a sufficiently diverse set of real-world errors remains challenging \cite{krishnan2017boostclean}. In this sense, it is natural to come up with the idea that we can manually generate errors for the model training.

Existing error generation methods primarily rely on rule-based approaches, with BART~\cite{arocena2015messing} being the most representative framework. BART introduces errors into clean datasets by applying predefined constraints, such as functional dependencies (FDs) \cite{mandros2017discovering,fan2008conditional} and denial constraints (DCs)~\cite{bian2024discovering,pena2022fast,chu2013discovering}, ensuring that the generated errors are systematically detectable. By enforcing a controlled corruption process, BART provides a benchmark for evaluating data cleaning algorithms.


\begin{figure}[t]
\centering
\setlength{\fboxrule}{0pt}
\fbox{\includegraphics[width=0.46\textwidth]{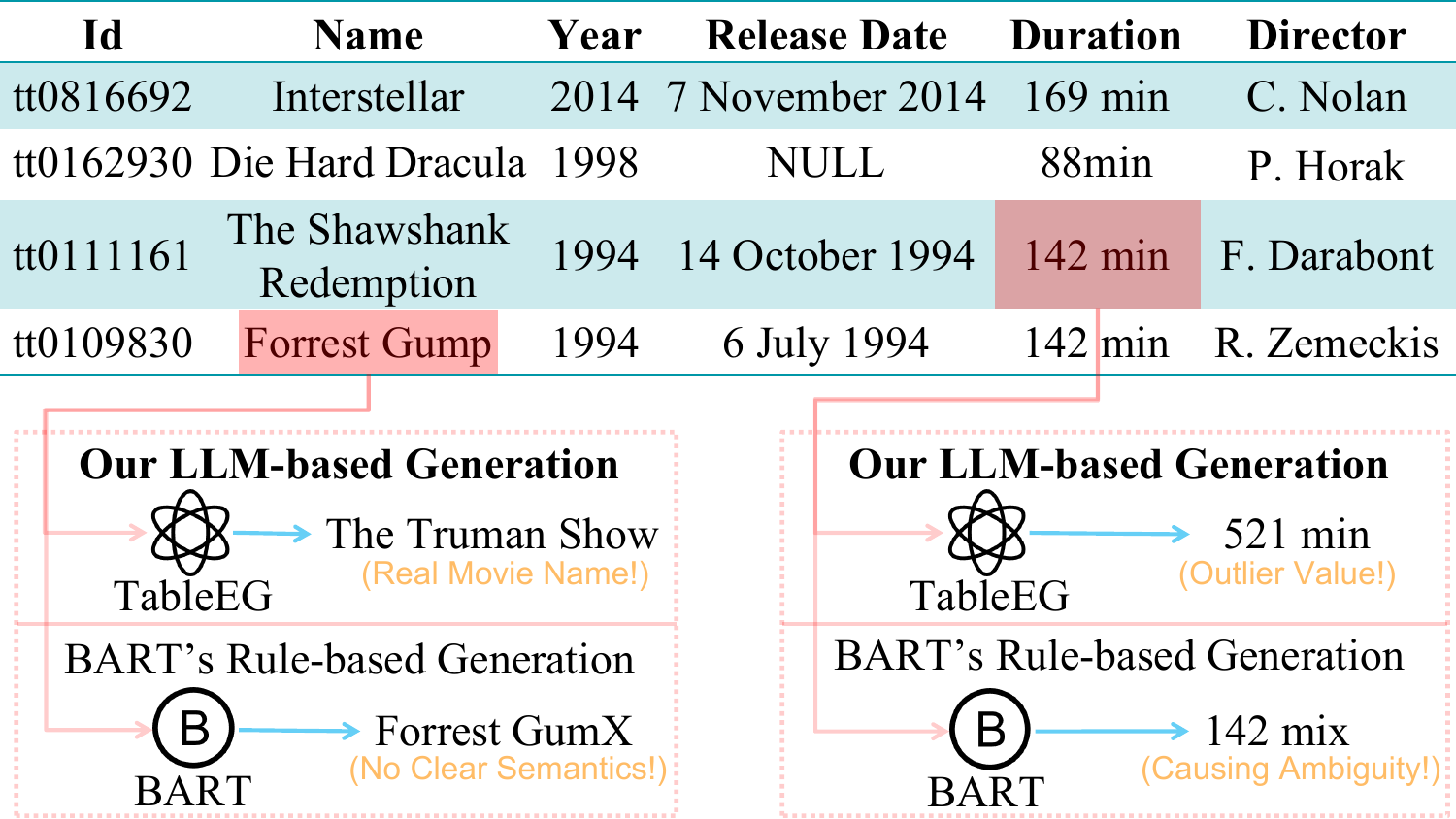}}
\caption{An example of the generated errors by BART \cite{arocena2015messing} and our TableEG, over the Movie dataset.}
\label{fig:compare-BART-TableEG}
\end{figure}

However, BART faces two major limitations that hinder its applicability in real-world scenarios. First, its generated errors are confined to predefined patterns (\emph{e.g.,} inserting or deleting characters within a word, or randomly replacing a value in the same attribute), making it difficult to capture the complexity of genuine data errors. As illustrated in Figure~\ref{fig:compare-BART-TableEG}, to inject errors into the Movie dataset, BART’s rule-based generation may transform the movie title “Forrest Gump” into “Forrest GumX,” or the duration “142 min” into “142 mix”. Clearly, such generated values rarely occur in real-world errors. That is, its outputs lack diversity and fail to represent the genuine characteristics of erroneous data. Therefore, their significance for both training error detection models and serving as a benchmark is limited, considering that BART cannot reproduce the nuanced patterns and statistical properties of authentic data corruption.

Second, BART focuses predominantly on constraint-violating errors. Although BART is designed to support a variety of random errors, including typos, duplicated values, outliers, and even missing or bogus values, its rule-based mechanism is inherently limited. In practice, BART struggles to generate missing values~\cite{krishnan2017boostclean} that faithfully reflect the nuanced patterns present in real-world datasets and to produce domain-specific semantic inconsistencies~\cite{suchanek2011paris}.
For instance, in the Flights~\cite{rekatsinas2017holoclean} dataset, the attribute \textit{act\_arr\_time} may contain the value “Not available,” indicating missing or ambiguous data rather than a simple constraint violation. However, BART cannot generate such errors as it assumes all errors stem from various constraint violations. This limitation also weakens its suitability for benchmarking error detection algorithms, which must handle a broader and more diverse range of real-world data corruptions.

Given these limitations, it would be more reasonable to generate errors, which could reflect the diversified characteristics of real-world errors, including various types and semantics. Thus, in this study, we consider \textbf{generating authentic errors}, which are expected to be as close as possible to those potential errors in real scenarios, to better support data cleaning tasks and enhance the applicability of error detection and correction techniques.

Recent advancements in Large Language Models (LLMs) have achieved remarkable success in natural language processing (NLP) tasks, drawing increasing attention to their potential for structured tabular data processing tasks~\cite{li2024table,wu2024tablebench,zhang2023tablellama}. In this sense, it is intuitive to leverage LLMs' robust semantic comprehension and expansive knowledge to systematically generate authentic errors. However, unlike natural language, relational data requires maintaining two-dimensional structural integrity while accounting for inter-cell dependencies (\emph{e.g.,} primary-foreign key relationships) and domain-specific constraints (\emph{e.g.,} financial transaction rules). This raises the fundamental questions: \textit{Can LLMs generate contextually meaningful errors rather than entirely random or nonsensical corruptions?  Do these synthetic errors faithfully replicate the diverse real-world errors so that they can be reliably detected by current error detection algorithms?}

\subsection{Challenges}

Actually, as shown in our empirical study,\footnote{Please see the experimental results in Figure~\ref{fig:bar}, Figure~\ref{fig:k} and Table~\ref{tab:Jw_DJS} in Section \ref{sec:experiments}.} directly using Large Language Models (LLMs) to generate errors in tabular data is not effective enough, primarily due to two major challenges.

\label{sec:challenge1}
(1) LLMs exhibit limited awareness of two-dimensional table structures, making it difficult to flexibly handle inter-row and inter-column dependencies. For instance, operations such as merging or splitting cells and referencing data across columns are not as straightforward for LLMs as processing linear text, resulting in relatively simplistic error patterns and occasionally producing illogical substitutions across rows and columns. 

\label{sec:challenge2}
(2) To be practically useful, generated errors must align with real-world error distributions and encompass a wide range of error types—constraint violations, missing values, format inconsistencies, and more complex semantic conflicts. However, direct invocation of LLMs typically only yields superficial value modifications, failing to ensure that the distribution of generated errors reflects genuine data corruption patterns.

\subsection{Solution}

To tackle the challenges outlined in Section~\ref{sec:challenge2}, we design an instruction-tuned approach to enhance the ability of LLMs in generating realistic errors in tabular data. Our solution consists of two key components: instruction fine-tuning with real error annotations and task augmentation for structural awareness.

(1) \textbf{Instruction Fine-Tuning for Error Generation}. 
Instead of directly prompting LLMs to generate erroneous tabular data, we fine-tune the model with explicit cell-level error annotations derived from real-world datasets. Each training instance is structured with an instruction, an input table, and an output, where the instruction provides a high-level directive, the input table contains clean data, and the output introduces annotated erroneous cells. By incorporating these explicit supervisory signals, our approach enables the model to learn the distribution of real-world errors while filtering out unrealistic noise.

(2) \textbf{Task Augmentation for Structural Awareness. }
To enhance the model’s understanding of tabular structures, we formulate a diverse set of task-specific instruction templates that reinforce the relationships between rows, columns, and constraint dependencies. These tasks cover a broad spectrum of error types, including constraint violations, missing values, format inconsistencies, and semantic conflicts. By systematically exposing the model to structural variations and dependency constraints, we significantly improve its ability to generate errors that align with real-world corruption patterns.

Compared to directly using LLM for error generation, our instruction-tuned method significantly enhances the realism and structural fidelity of generated errors. By incorporating explicit supervision and task augmentation, it enables LLMs to capture complex inter-row and inter-column dependencies, addressing their inherent limitations in processing two-dimensional data. Moreover, the generated errors closely align with real-world distributions, encompassing a diverse range of error types beyond superficial value modifications. 

As shown in Figure~\ref{fig:compare-BART-TableEG}, the error generation model TableEG, enhanced through instruction fine-tuning and multi-task learning, identifies row and column structures in tabular data. Leveraging the memory and reasoning capabilities of LLMs, it retrieves semantically meaningful real-world errors from the knowledge base as substitutions—for instance, replacing a movie title in a record with an actual film name, such as "The Truman Show."  

Moreover, TableEG is not limited to generating constraint-violating errors; it also produces errors spanning outliers, missing values, and pattern violations (see Section~\ref{sec:errortypes} for classification details). Figure~\ref{fig:compare-BART-TableEG} illustrates an example of TableEG generating an outlier, \emph{i.e.,} ``521 min'', demonstrating that an instruction-tuned, multi-task enhanced LLM exhibits strong multi-category error generation capabilities, effectively filling the gaps left by BART.

\subsection{Contributions}
\label{sect-contributions}

Our work makes the following key contributions:

(1) We devise a formulation and a structured framework for error generation in tabular data based on LLMs (see Section~\ref{sec:framework}). Our approach is built upon 12 real-world datasets spanning 10 diverse domains, ensuring that the training set reflects varied error types and semantics in real applications. 

(2) We introduce TableEG, a model that extends traditional table-task fine-tuning by incorporating three error-related sub-tasks (detailed in Section~\ref{sec:methodology}) Leveraging manually labeled error data from our standardized annotation collection, we synthesize triplets $(I, T, O)$ that serve as training examples for our TableEG. Furthermore, TableEG supports configurable error ratios and types, enabling the production of domain-relevant and distributionally aligned synthetic errors.

(3) We design a comprehensive set of evaluation strategy in Section~\ref{sec:evaluationstrategy} to measure how closely the generated errors align with real-world error. Our framework quantifies both pattern fidelity and distributional consistency, providing objective metrics that validate the realism and practical utility of our synthetic errors.

(4) Extensive experiments on seen and unseen datasets demonstrate that our TableEG model outperforms existing error generation methods in terms of pattern and distribution similarity. These methods include the rule-based BART and the deep learning-based GPT-3.5 (Turbo) without fine-tuning. 
Moreover, various error detection algorithms exhibit comparable performance on both LLM-generated and real-world errors, further confirming that our approach successfully mirrors genuine data corruption characteristics. (See Section~\ref{sec:experiments} for details.)


\section{Preliminaries}
\label{sec:preliminaries}

In this section, we lay the groundwork for our study by first defining common data errors in Section~\ref{sec:errortypes} and then discussing the challenges of applying Large Language Models (LLMs) to tabular data in Section~\ref{sec:largemodelsfortabletask}. In particular, we provide a formal definition for data errors (Definition~\ref{def:dataerror}) and categorize data errors into four types. We further review the limitations of standard LLMs in handling the inherent two-dimensional structure and inter-cell dependencies of tables. These discussions motivate our introduction of a triplet-based representation for modeling table tasks, which serves as a foundation for our proposed framework.

\subsection{Types of Data Errors}
\label{sec:errortypes}

In data cleaning and quality research, errors are typically defined as deviations of data values from their ground truth~\cite{abedjan2016detecting}. 

\begin{definition}[Data Error]
\label{def:dataerror}
Consider a relational dataset $D$ under a schema $\mathcal{R}$ with attributes 
$\{A_1, A_2, \dots, A_m\}$. Let $D^*$ be the cleansed version of $D$, reflecting 
the ground truth. For any tuple $t_i \in D$ and any attribute $A_j \in \mathcal{R}$, 
if $t_i[A_j] \neq t_i^*[A_j]$, then the cell $t_i[A_j]$ is defined as a 
\emph{data error}. 
\end{definition}

These errors can arise from various reasons—such as noise in data acquisition, manual entry mistakes, rule violations, or inconsistencies introduced during data integration. Several studies have proposed classification schemes to categorize these errors either based on their origins~\cite{rahm2000data,kim2003taxonomy,hellerstein2013quantitative} or on the violations of constraints and patterns observed in structured data~\cite{ilyas2015trends}.


Our classification scheme is broadly consistent with the taxonomy proposed by Abedjan et al.~\cite{abedjan2016detecting}, which, although not exhaustive, serves as a well-established framework for systematically categorizing errors commonly identified in real-world datasets.  However, our classification differs in several key aspects. First, we explicitly separate missing values as an independent category, acknowledging their prevalence in real-world data. Second, we incorporate duplicates within the broader category of rule violations. Lastly, our definition of rule violations extends to any value that violates a form of denial constraints, encompassing a wider range of integrity violations. Consequently, we categorize errors into four primary types—outliers, missing values, rule violations, and pattern violations—capturing the most prevalent and impactful error types observed in real-world tabular datasets.

\textbf{Outliers}: Outliers are data values that significantly deviate from the expected distribution within a column, either numerically or categorically. They often stem from measurement errors, manual entry mistakes, or data integration inconsistencies. For example, an unusually high transaction amount in financial records may indicate an anomaly or an input error. 
    
\textbf{Missing Values}: Missing values occur when data cells are empty, improperly filled, or lost due to format conversion issues or inconsistent data integration. They compromise data completeness and can introduce biases in downstream analysis. They often represented as NULL values, empty strings, or placeholder entries (\emph{e.g.,} “N/A”, “9999-999999”).


\textbf{Rule Violations}: We employ a set of predefined integrity constraints, primarily in the form of DCs~\cite{chu2013discovering}), FDs~\cite{mandros2017discovering} and CFDs~\cite{fan2008conditional} to safeguard data consistency. Any record that fails to meet these constraints is classified as a rule violation. Such violations often involve contradictory, prohibited, or otherwise infeasible values, thereby severely undermining data integrity. They typically arise from data entry errors, integration inconsistencies, or improper transformations, and are commonly detected through constraint validation or rule-based inference.

\textbf{Pattern Violations}: Pattern violations refer to data values that fail to adhere to expected syntax, structure, or semantic requirements. Syntax errors encompass format mismatches (\emph{e.g.,} “10-15” vs. “October 15th”) and typographical errors. Semantic inconsistencies arise when values conflict with established domain constraints (\emph{e.g.,} a ZIP code that does not match its corresponding city). These violations will significantly hinder data parsing, querying, and subsequent processing tasks.
    

\begin{figure*}[t!]
\centering
\setlength{\fboxrule}{0pt}
\fbox{\includegraphics[width=0.9\textwidth]{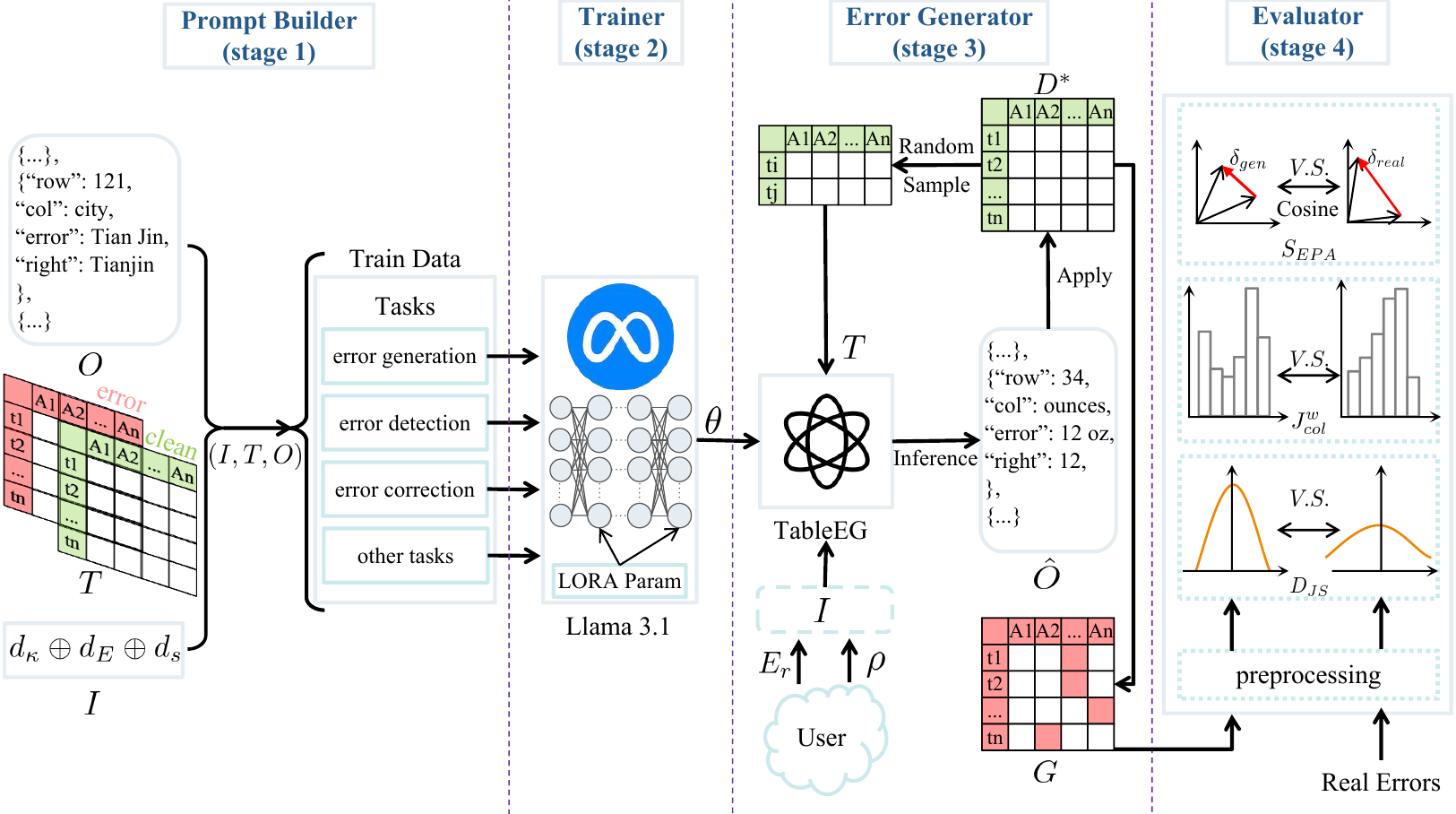}}
\caption{Overview of training and utilizing our TableEG for error generation.}
\label{fig:pipeline}
\end{figure*}

\subsection{LLMs for Table Task}
\label{sec:largemodelsfortabletask}


LLMs have attracted many researchers with their large-scale parameters and excellent performance on various text processing tasks. However, their application to structured tabular data presents unique challenges due to their lack of inherent awareness of table-specific constraints. In the following sections, we first review recent advancements in LLMs and discuss their limitations in handling tabular data. Finally, we explore how instruction-tuned approaches can enhance LLMs by explicitly integrating table-specific constraints to improve their performance in table-related tasks.

\subsubsection{Recent Advancements in Large Language Models (LLMs)}
\label{sec:llms}

Recent advancements in Large Language Models (LLMs) have demonstrated remarkable capabilities in natural language understanding and generation, primarily trained on vast corpora of sequential text~\cite{vaswani2017attention,brown2020language}. However, directly applying LLMs to relational data and tabular tasks remains nontrivial due to the inherent structural differences between natural language and relational tables. Unlike textual data, which follows a linear sequence, tabular data is inherently two-dimensional, often governed by schema constraints, cross-column dependencies, and entity relationships~\cite{dong2024large}. These structural properties impose significant challenges when leveraging LLMs for table-based tasks such as error generation, detection, and repair.

\subsubsection{Challenges of Applying LLMs to Tabular Data}
\label{sec:challenges}

First, traditional LLMs primarily operate under a token-based autoregressive or masked prediction framework, which excels in modeling linguistic dependencies but struggles with structured tabular constraints~\cite{yin2020tabert}. In relational tables, data integrity often depends on multi-column dependencies, such as functional dependencies (FDs) or domain constraints, which LLMs do not inherently recognize~\cite{chu2016data}. Consequently, naive applications of LLMs to table-based tasks often result in syntactically plausible yet semantically invalid outputs, where column dependencies are violated, inconsistent values are introduced, or key constraints are broken. 
Second, LLMs tend to generate errors based on linguistic priors rather than data-specific corruption patterns. Unlike real-world data errors—which may arise from user input mistakes, OCR misrecognition, or systematic extraction faults—LLM-generated corruptions often exhibit artificial substitution biases that fail to reflect natural error distributions. This discrepancy limits the effectiveness of LLM-generated datasets for evaluating error detection and repair models, as the synthetic errors may not align with the challenges encountered in real-world applications.

\subsubsection{Leveraging Instruction-Tuned LLMs for Table Error Processing}
\label{sec:instruction-tuning}

To address these limitations, recent studies have explored instruction-tuned LLMs that explicitly integrate table-specific constraints and domain knowledge into their fine-tuning process~\cite{li2024table}. Our approach builds on this paradigm by incorporating structural metadata—such as schema constraints and functional dependencies—into the LLM's training pipeline. By embedding such constraints within the instruction design, we guide the model to generate contextually valid errors that better reflect real-world data corruptions while maintaining table consistency.

\section{Framework Overview}
\label{sec:framework}

In this section, we lay the foundation for our work by first formally defining the error generation problem in Section~\ref{sec:definitionEGT} and then presenting our structured framework for addressing it. In Section~\ref{sec:overview}, We thoroughly describe each step in training and applying TableEG for error generation. Finally, in Section~\ref{sec:datasets}, we detail the 12 real-world datasets from 10 diverse domains used in our study, ensuring that the training data encapsulates authentic and varied error types. These subsections collectively provide a solid foundation for our approach and pave the way for the detailed methodology and experimental evaluations that follow.

\subsection{Problem Statement}



To fully appreciate the challenges in error generation for tabular data and to establish a foundation for our approach, it is critical to formally define the error generation task. In our context, generating realistic errors is not merely about altering data randomly; it requires producing modifications that accurately mirror the error patterns and distributions observed in real-world datasets. This formalization guides the design of our training strategy.

\label{sec:definitionEGT}
\begin{problem}
Given an original dataset $D^*$, the objective is to generate a modified dataset $G$ by introducing synthetic noise that closely mimics real-world data errors. The generated dataset $G$ should not only replicate the diverse error types observed in real-world scenarios but also preserve the underlying structural patterns and statistical distributions of the original data.
\end{problem}

Our primary goal is to enable Large Language Models (LLMs) to generate highly realistic tabular errors by exploiting specialized fine-tuning strategies that effectively reflect the inherent two-dimensional structure of tables. Given a clean table \(D^*\) containing \(n\) rows and \(m\) columns, our task is to learn an error generation function \(f_{\theta}\) via LLMs, thereby producing a modified table \(G\) that convincingly simulates realistic data corruption.

To achieve this, the clean dataset \(D^*\) is partitioned and processed by our TableEG framework, which applies \(f_{\theta}\) to generate \(G\) with diverse and contextually plausible error patterns. TableEG systematically learns to inject errors by considering inter-row and inter-column dependencies, ensuring that generated corruptions align with real-world error characteristics. 

In the following section, we provide a comprehensive overview of our error generation framework, detailing the construction process of instruction-driven data, the specifics of error understanding tasks, and the comprehensive metrics used to evaluate the quality of generated errors.


\subsection{TableEG Overview}
\label{sec:overview}

As shown in Figure~\ref{fig:pipeline}, we provide a detailed description of each step involved in training and utilizing TableEG for error generation. The process spans from transforming raw dirty datasets into training prompts to fine-tuning the model, generating realistic synthetic errors, and ultimately evaluating their authenticity.

\noindent
\textbf{Stage 1: Prompt Builder.} Our approach begins by ingesting both clean and dirty data from real-world datasets, which are manually annotated to serve as the output labels for the model. Next, we assign different task settings to distinct generators that produce corresponding instructions. Leveraging our triplet representation \((I, T, O)\), the Prompt Builder constructs the instruction \(I\) by integrating three key components: Task Description (\(d_\kappa\)), Error Type Description (\(d_E\)) and Contextual Suffix (\(d_s\)). We detailed it into Section~\ref{sec:tabletasksummary} and Section~\ref{sec:promptbuilder}.

\noindent
\textbf{Stage 2: Trainer.} 
In this stage, the training set generated in Stage 1 is fed into a pretrained Large Language Model, LLaMA3.1-8B, which is then fine-tuned using a LoRA-based low-rank adaptation technique. This fine-tuning process yields TableEG, a model specifically optimized for generating realistic errors in tabular data. We carefully adjust the training parameters to ensure that the model effectively learns to capture the structural and contextual nuances of table errors. Detailed training parameters and settings are provided in Section~\ref{sec:trainingprocedure}.

\noindent
\textbf{Stage 3: Error Generator.}
TableEG demonstrates the ability to generate realistic errors in tabular data. Given a clean dataset \(D^*\) and user-specified error parameters (error type distribution ratio \(E_r\) and error ratio \(\rho\)), our framework first produces a task-specific instruction \(I\) that reflects the desired error characteristics. Simultaneously, a sub-table \(T_i\) is randomly sampled from \(D^*\), to capture representative structural contexts. The fine-tuned model then performs inference on this input, selecting appropriate cell locations for error injection and outputting a corresponding modification \(\hat{O}\). Finally, \(\hat{O}\) is applied to the clean table \(D^*\) to produce the modified (or “dirty”) table \(G\), which exhibits synthetic errors that mimic real-world error distributions. Detailed algorithmic procedures for this stage are provided in Section~\ref{sec:errorgenerator}.

\noindent
\textbf{Stage 4: Evaluator. }
To rigorously evaluate TableEG’s generation capability, we conduct a series of quantitative and qualitative experiments to assess how closely its generated errors approximate real-world errors.  
First, we define $S_{EPA}$ to measure pattern alignment between generated and real errors (details in Algorithm~\ref{alg:evaluation_generated_error}). 
Secondly, we employ weighted Jaccard Similarity ($J^w_{\text{col}}$) and Jensen-Shannon Divergence ($D_{JS}$) to quantify the similarity of column-wise error distributions. 
Furthermore, we conduct qualitative auxiliary experiments by applying error detection algorithms separately to real errors and TableEG-generated errors on the same datasets. By analyzing whether detection performance varies significantly between the two, we further validate that TableEG can simulate authentic data inconsistencies that closely resemble those found in production databases. Our evaluation strategy is detailed in Section~\ref{sec:evaluationstrategy}.

\subsection{Datasets}
\label{sec:datasets}

In this section, we describe the datasets used for training, selected based on the following key considerations:

(1) \textbf{Authenticity}: Since our goal is to generate errors that closely resemble real-world data corruption, we prioritize datasets containing naturally occurring errors rather than artificially synthesized noise. This ensures that the generated errors reflect realistic patterns and distributions.

(2) \textbf{Domain Diversity}: To enhance the model’s generalization
ability, the datasets span diverse domains. This exposure allows the model to learn varied tabular structures and semantic relationships, improving its adaptability to unseen data.

(3) \textbf{Error Type Diversity}: The datasets are selected to encompass the four primary error types mentioned in Section~\ref{sec:errortypes}. They include a variety of error manifestations, allowing the model to learn distinct error generation mechanisms based on data pattern rather than being constrained by specific datasets.

By ensuring a balanced combination of these factors, we aim to build a model capable of generating realistic and diverse errors applicable to a broad range of tabular data. The datasets used in this study are primarily sourced from those introduced in \cite{li2021cleanml, mahdavi2019raha}, which are primarily designed for error detection and data cleaning tasks.

\begin{table}[t]
  \caption{Dataset characteristics. The error types include missing value (\textbf{M}), pattern violation (\textbf{P}), rule violation (\textbf{R}), and outliers (\textbf{O}).}
  \small
  \label{tab:dataset_characteristics}
  \centering
  \setlength\tabcolsep{2.8pt}
  \begin{tabular}{l c c c c l}
    \toprule
    Dataset & $|D|$ & $|R|$ & {Domain} & {Error Rate} & {Type} \\
    \midrule
    Rayyan & 1,000 & 11 & Academic & 8.62\% & M, P, R \\
    Company & 128,889 & 7 & Business & 34.21\% & P, R \\
    Marketing & 8,993 & 14 & Business & 21.29\% & M, P \\
    Movie (Metadata) & 7,390 & 17 & Entertainment & 6.10\% & M, P \\
    Movie (BoxOffice) & 9,329 & 7 & Entertainment & 7.31\% & P \\
    Credit & 150,000 & 10 & Finance & 2.33\% & M, O \\
    Beers & 2,410 & 11 & Food & 12.66\% & M, P, R \\
    Restaurant & 12,007 & 10 & Food & 0.53\% & P \\
    Hospital & 1,000 & 20 & Health & 2.55\% & P, R \\
    Airbnb & 42,492 & 40 & Hospitality & 0.22\% & M, O \\
    University & 286 & 17 & Education & 13.97\% & P \\
    Sensor & 62,076 & 8 & Technology & 0.01\% & O \\
    Flights & 2,376 & 7 & Transportation & 24.15\% & M, P, R \\
    \bottomrule
  \end{tabular}
\end{table}

\textbf{Rayyan:} This dataset contains 1,000 academic references extracted from systematic review databases such as MEDLINE and EMBASE. It includes bibliographic attributes like article title, journal name, ISSN, language, and author list. Due to the integration of multiple sources and manual data entry, the dataset contains missing values (M), pattern violations (P), and rule violations (R). These errors often arise from inconsistencies in journal abbreviations, misplaced author names, and incorrect citations.

\textbf{Company:} This dataset consists of 128,889 business records, sampled from a large corpus of company-related data. Each record includes attributes such as company name, country, industry sector, and financial indicators. The dataset exhibits a high error rate (34.21\%), with pattern violations (P) due to inconsistent company name formatting and rule violations (R) caused by duplicate and conflicting records.

\textbf{Marketing:} Containing 8,993 records from a household income survey, this dataset includes 14 demographic attributes such as sex, education, and employment status. The dataset has missing values (M) due to incomplete survey responses and pattern violations (P) stemming from inconsistent data formats.

\textbf{Movies (Metadata):} This dataset sourced from the Magellan repository, containing 7,390 movie records with attributes like title, year, director, cast, language, genre, and rating. It is used for entity matching and deduplication, as duplicate entries exhibit inconsistencies in naming, genre classification, and language format, resulting in pattern violations (P). Additionally, missing values (M) appear in metadata fields such as filming locations.

\textbf{Movies (Box Office):} This dataset consists of 9,329 movie records from IMDB and TMDB, focusing on genre classification and sentiment analysis. It contains attributes like title, genre, budget, language, and score but suffers from inconsistent genre labels, duplicate records, and missing values (M) in budget and language fields, affecting predictive modeling.

\textbf{Credit:} A financial dataset containing 150,000 credit records with attributes such as income, age, and number of dependents. The dataset is affected by missing values (M) in financial attributes and numerical outliers (O) due to extreme income values.

\textbf{Beers:} This dataset comprises 2,410 beer-related records from brewery databases, including information on brand, alcohol percentage, and taste profile. Errors include missing values (M), pattern violations (P), and rule violations (R) due to inconsistencies in alcohol content reporting.

\textbf{Restaurant:} The dataset includes 12,007 restaurant listings sourced from Yelp and Yellowpages, with attributes like location, category, and customer rating. Pattern violations (P) occur due to discrepancies in restaurant names and price range formatting.

\textbf{Hospital:} A healthcare dataset with 1,000 hospital records containing 20 attributes, such as facility name, location, and specialization. The dataset includes pattern violations (P) and rule violations (R) caused by inconsistent hospital categorizations and conflicting information about specialties.

\textbf{Airbnb:} This dataset consists of 42,492 rental property listings, containing 40 attributes such as location, price, and host information. Missing values (M) are prevalent in availability dates, and numerical outliers (O) appear in rental prices.

\textbf{University:} A small dataset with 286 university records, each containing 17 attributes, including SAT scores, tuition fees, and institution type. Pattern violations (P) arise from inconsistent representations of state abbreviations and rankings.

\textbf{Sensor:} This dataset includes 62,076 sensor readings collected from IoT devices, with attributes such as temperature, humidity, and light intensity. The dataset contains numerical outliers (O) due to extreme fluctuations in environmental readings.

\textbf{Flights:} This dataset consists of 2,376 flight records with attributes like departure time, airline, and ticket price. Errors include missing values (M), pattern violations (P), and rule violations (R), arising from inconsistencies in time zone processing and fare calculations.

\begin{table*}[htbp]
\centering
\caption{A summary of our table tasks.}
\label{tab:tabletasks}
 \begin{threeparttable}   
\begin{tabular}{|c|c|c|c|c|}
\hline
\textbf{Task id} & \textbf{Category}                & \textbf{Task name}      & \textbf{Task description}                                                                                                  & \textbf{Usage}              \\ \hline
Task-1\label{T1}               & \multirow{3}{*}{Error-related}   & Error Generation        & \begin{tabular}[c]{@{}c@{}}Select a cell in the given table and modify its value \\ to introduce an error.\end{tabular}    & Train \& Test               \\ \cline{1-1} \cline{3-5} 
Task-2               &                                  & Error Detection         & Detect a erroneous cell in a given table                                                                                   & \multirow{7}{*}{Train only} \\ \cline{1-1} \cline{3-4}
Task-3               &                                  & Error Correction        & \begin{tabular}[c]{@{}c@{}}Select a cell in the given table and modify its value \\ to repair an error\end{tabular}        &                             \\ \cline{1-4}
Task-4               & \multirow{5}{*}{Error-unrelated$^*$} & Row/Column Augmentation & \begin{tabular}[c]{@{}c@{}}Augment the table with additional rows/columns\\  compatible with the given table.\end{tabular} &                             \\ \cline{1-1} \cline{3-4}
Task-5               &                                  & Row/Column Swapping     & \begin{tabular}[c]{@{}c@{}}Swap the positions of specified rows/columns \\ in the given table.\end{tabular}                &                             \\ \cline{1-1} \cline{3-4}
Task-6               &                                  & Row/Column Filtering    & Filter specified rows/columns in the given table.                                                                          &                             \\ \cline{1-1} \cline{3-4}
Task-7               &                                  & Header Matching         & \begin{tabular}[c]{@{}c@{}}Match column headers with extracted data values \\ in the same table.\end{tabular}              &                             \\ \cline{1-1} \cline{3-4}
Task-8               &                                  & Table Summarization     & Generate a summary based on the table content.                                                                             &                             \\ \hline
\end{tabular}
  \begin{tablenotes}
  \item[*] The Error-unrelated tasks are inspired by Table-GPT~\cite{li2024table} and are designed to enhance the model's ability to recognize row and column positions while improving its overall understanding of tabular data.
  \end{tablenotes}
  \end{threeparttable}
\end{table*}

\section{Methodology}
\label{sec:methodology}

In this section, we detail our approaches. Building on the overview presented in Section~\ref{sec:overview}, we elaborate on the specific components of our framework as follows. In Section~\ref{sec:tabletasksummary}, we introduce the summary of our table task, which outlines our approach to handling table tasks with a focus on the three error-related tasks. In Section~\ref{sec:promptbuilder}, we detail the prompt builder module, which constructs complete training triplets \((I, T, O)\) by synthesizing task-specific instructions and corresponding input tables. Section~\ref{sec:trainingprocedure} describes our training procedure, including parameter settings and the fine-tuning process that yields the TableEG model. Finally, in Section~\ref{sec:errorgenerator}, we explain how our TableEG model is applied for error generation, enabling users to control error ratios and types to produce datasets for downstream tasks.

\subsection{Table Task Summary}
\label{sec:tabletasksummary}
In Section~\ref{sec:introduction}, we discuss that our primary objective is to enhance the ability of LLMs to generate realistic errors. To this end, we carefully select a set of tasks for fine-tuning the model. Drawing inspiration from previous works~\cite{li2024table, zhang2023tablellama}, we first incorporate a series of error-agnostic tasks to bolster the basic reading and writing capabilities of LLMs on tabular data. These tasks enable the model to analyze two-dimensional tables and accurately identify specific rows and columns. For instance, as illustrated in Table~\ref{tab:tabletasks}, Row/Column Augmentation (Task-4) improves the model’s understasnding of row and column identifiers, Header Matching (Task-7), Row/Column Swapping (Task-5), and Filtering (Task-6) enhance its ability to read positional information, and Table Summarization (Task-8) reinforces its comprehension of table content.

Simultaneously, we design three error-related tasks that directly address the challenge of error generation in tabular data. We devise a structured framework composed of three interrelated sub-tasks: Error Generation (Task-1), Error Detection (Task-2), and Error Correction (Task-3). Notably, Error Generation (Task-1) and Error Correction (Task-3) form a coupled pair, with Error Detection (Task-2) serving as a prerequisite for effective correction. This framework leverages the intrinsic relationships among these tasks to enable comprehensive error handling, thereby improving the overall robustness of LLMs in managing errors.

\label{task:errorgeneration}
\textbf{Error Generation (EGT)} : the model is given a clean table and is required to introduce realistic errors into it, mimicking common issues found in real-world datasets. The input table $T_{\text{clean}}$ contains correct values without any injected noise, while the output $O_{\text{dirty}}$ consists of structured annotations specifying the error location, the type of error, and the generated incorrect value. The instruction $I_{\text{EGT}}$ consists of a high-level task description along with an error type description, allowing the model to either generate a variety of errors or focus on specific types, such as outliers, missing values, rule violations, or pattern violations. The model learns to generate plausible erroneous values while ensuring consistency with the dataset’s structure and statistical properties. The overall task formulation is:

\begin{equation}
EGT(T) = (I^{EGT}, T_{\text{clean}}, O_{\text{dirty}})
\end{equation}

\textbf{Error Detection (EDT)}: This task requires the model to identify erroneous entries in a table containing pre-introduced errors. The input, $T_{\text{dirty}}$, consists of data with injected errors, simulating real-world inconsistencies. The expected output, $O_{\text{detection}}$, provides structured annotations specifying the erroneous cells and classifying the corresponding error types. The instruction, $I_{\text{EDT}}$, guides the model in systematically analyzing the table to determine the locations and categories of errors. A key challenge in this task is ensuring the model’s ability to generalize across diverse error distributions, requiring it to effectively differentiate between natural variations in data and actual inconsistencies.

\begin{equation}
EDT(T) = (I^{EDT}, T_{\text{dirty}}, O_{\text{detection}})
\end{equation}

\textbf{Error Correction (ECT)} : This task focuses on recovering the correct values for erroneous cells detected within a dataset. Given an input table $T_{\text{dirty}}$ that contains errors, the model is expected to infer the correct values and produce them as part of the output $O_{\text{correction}}$. The expected output includes structured annotations that specify the erroneous cells alongside the predicted correct values. The instruction $I_{\text{ECT}}$ directs the model to correct errors based on contextual clues and logical consistency within the dataset. Unlike error detection, which only identifies issues, this task requires the model to leverage statistical reasoning, domain-specific patterns, and implicit constraints to restore data accuracy.

\begin{equation}
ECT(T) = (I^{ECT}, T_{\text{dirty}}, O_{\text{correction}})
\end{equation}

This triplet representation offers a standardized framework for defining table-related tasks, facilitating LLMs' ability to generalize across multiple table-based operations, including table completion, error detection, and error correction.

While our primary focus is the error generation, error detection and correction serve as essential auxiliary tasks that improve the model’s ability to learn meaningful error patterns. Error Detection provides an intermediate supervisory signal, helping the model differentiate between true errors (\emph{e.g.,} typos, rule violations) and clean cells while Error Correction reinforces contextual reasoning by requiring the model to restore accurate values based on learned patterns. By jointly manipulate these interrelated tasks, our framework enhances the LLM’s ability to not only generate realistic errors but also understand the distribution and properties of real-world tabular errors, ultimately improving its generalization to unseen data scenarios.

\subsection{Prompt Builder}
\label{sec:promptbuilder}

As highlighted in \emph{Challenge~1} (Section~\ref{sec:challenge1}) of our introduction, directly applying LLMs to table tasks and error generation faces significant difficulties. Traditional NLP models primarily handle sequential text, but relational data imposes two-dimensional structural constraints and inter-cell dependencies (\emph{e.g.,} referencing multiple columns or rows). Consequently, LLMs often struggle to flexibly merge, split, or substitute cells while preserving logical consistency across columns and rows. Moreover, the simplistic substitution patterns that LLMs naturally produce can fail to reflect real-world error distributions.

To address the aforementioned challenges in table-tuning and error-related tasks, we build on instruction fine-tuning~\cite{ouyang2022training} and the table-tuning paradigm introduced by Li \emph{et al.}~\cite{li2024table}. Specifically, we design a triplet representation $(I, T, O)$ to tackle error-related tasks in tabular data.

\begin{definition}[Triplet Representation]
\label{def:triplet}
Each table-based task is modeled as a triplet $(I, T, O)$, where $I$ is an instruction specifying the task goal, $T$ is the (sampled) input table, and $O$ is the expected structured output.
\end{definition}

Building on the triplet representation $(I, T, O)$ introduced in Definition~\ref{def:triplet}, 
Algorithm~\ref{alg:table_triplet_generation} generates these triplets from a manually annotated collection $\mathcal{J}$ of JSON records. $\mathcal{J}$ is derived from a clean, two-dimensional table dataset \(D\) and contains records that have been manually annotated. Each record \(t \in \mathcal{J}\) is represented as an object that comprises the original table data along with an annotation field \(t.E\) (\emph{i.e.,} the error type \(E\)).

\begin{algorithm}[ht]
\caption{Generating Triplets for LLM table-tuning}
\label{alg:table_triplet_generation}

\begin{algorithmic}[1]
    \Statex \textbf{Input}: Dataset $D$, error type $E$, table task \(\kappa\), annotation $\mathcal{J}$
    \Statex \textbf{Output}: A synthesized dataset $\mathbb{D}_{\text{train}} = \{(I, T, O)\}$
    \State $\mathbb{D}_{\text{train}} \gets \emptyset$
    \For{each $j \in \mathcal{J}$}
        \State \label{alg1line:getE}$E \gets j.E$ 
        \State \label{alg1line:getO}$O \gets j$ \Comment{Select labeled error annotations}
        \State \label{alg1line:getI}$I \gets \text{Generate-Instruction}(\textit{$\kappa$}, E)$ \Comment{Algorithm~\ref{alg:generate_instruction}}
        \State\label{alg1line:getT} $T \gets \text{Construct-Input-Table}(D,j)$ \Comment{Algorithm~\ref{alg:sampling_table_construction}}
        \State \label{alg1line:getA}$\mathbb{D}_{\text{train}} \gets \mathbb{D}_{\text{train}} \cup \{(I, T, O)\}$
    \EndFor
    \State \textbf{return} $\mathbb{D}_{\text{train}}$
\end{algorithmic}
\end{algorithm}

We processe each tuple \(j \in \mathcal{J}\) sequentially. The table task is represented by $\kappa$, which specifies the overall task objective (\emph{e.g.,} error generation, detection, or correction). For each \(j \in \mathcal{J}\), it extracts the error type \(E\) from \(j.E\) (Line~\ref{alg1line:getE}) and invokes the function \(\text{Output}(t)\) to construct the structured output \(O\) from the annotated error details (Line~\ref{alg1line:getO}). Next, the subroutine \(\text{Generate-Instruction} (\kappa, E)\) (detailed in Algorithm~\ref{alg:generate_instruction}) dynamically composes a task-specific instruction \(I\) (Line~\ref{alg1line:getI}). Subsequently, the input table \(T\) is built by calling \(\text{Construct-Input-Table}(t)\) (see Algorithm~\ref{alg:sampling_table_construction}), which samples and merges relevant rows to preserve the table's structural context (Line~\ref{alg1line:getT}). Finally, the triplet \((I, T, O)\) is added to the output set \(\mathbb{D}_{\text{train}} \) (Line~\ref{alg1line:getA}), and after processing all records in \(\mathcal{J}\), \(\mathbb{D}_{\text{train}} \) is returned as the final output. This systematic procedure ensures that each triplet accurately encapsulates the task instruction, contextual table data, and corresponding error annotation, thereby providing a unified input-output representation for LLM fine-tuning on error-related table tasks. The overall algorithm is shown in Algorithm~\ref{alg:table_triplet_generation}.


After defining the triplet \((I, T, O)\) in Definition~\ref{def:triplet}, we explain how each component is derived from an annotated record. Specifically, we describe the process of constructing the task-specific instruction \(I\), the input table \(T\), and the structured output \(O\) through dedicated procedures.

\textbf{Instruction} ($I$): The instruction serves as a natural language directive that defines the task requirements and guides the model’s operations. We generate a task-specific instruction \(I\) by combining three components: a task description \(d_\kappa\), an error type description \(d_E\), and a contextual suffix \(d_s\). First, we retrieves \(d_\kappa\) (Line~\ref{alg2line:getTaskDesc}), which explains the overall objective of the task (\emph{e.g.,} “error generation”) based on the task indicator \(\kappa\). Next, it extracts \(d_E\) (Line~\ref{alg2line:getErrorDesc}) that adapts the instruction to the specific error type \(E\) provided by the annotation. Then, it obtains \(d_s\) (Line~\ref{alg2line:getSuffixDesc}) to supply additional details such as the expected output format. Finally, these three components are concatenated (Line~\ref{alg2line:Join}) to form the complete instruction \(I\), which is returned for use in the triplet generation process. The overall algorithm is shown in Algorithm~\ref{alg:generate_instruction}.


\begin{algorithm}[htbp]
\caption{Generate-Instruction($\kappa, E$)}
\label{alg:generate_instruction}
\begin{algorithmic}[1]
    \Statex \textbf{Input}: Table task \(\kappa\), error type \(E\)
    \Statex \textbf{Output}: Task-specific instruction \(I\)
    \State \label{alg2line:getTaskDesc} \(d_\kappa \gets\) \text{Retrieve-Task-Description}\((\kappa)\)
    \State \label{alg2line:getErrorDesc} \(d_E \gets\) \text{Retrieve-Error-Type-Description}\((E)\) 
    \State \label{alg2line:getSuffixDesc} \(d_s \gets\) \text{Retrieve-Suffix-Description} \((\kappa)\)
    \State \label{alg2line:Join} \(I \gets d_\kappa \oplus d_E \oplus d_s\) 
    \State \textbf{return} \(I\)  
\end{algorithmic}
\end{algorithm}

\textbf{Table} (\(T\)): The input table provides the structured data context for task execution. Due to token length limitation in LLMs input, it is impractical to feed an entire table into the model. Hence, we adopt a sampling-based strategy that constructs \(T\) by selecting a representative subset of rows. In our algorithm, let \(j^{(s)}\) denote the set of rows directly extracted from a given annotated record (Line~\ref{alg3line:ExtractTuplePairs}) and \(j^{(a)}\) denote additional rows sampled from the clean dataset \(D\) (Line~\ref{alg3line:SampleRows}) while excluding those already in \(j^{(s)}\). These two sets are then merged and shuffled to produce the final input table \(T\) (Line~\ref{alg3line:MergeandShuffle}), which is returned in a serialized Markdown format (\emph{i.e.,} as a Markdown table representation).

\begin{algorithm}[htbp]
\caption{Construct-Input-Table(\textit{D,t})}
\label{alg:sampling_table_construction}
\begin{algorithmic}[1]
    \Statex \textbf{Input}: Dataset \(D\), annotated record \(j\)
    \Statex \textbf{Output}: Constructed serialized table \(T\)
    \State \label{alg3line:ExtractTuplePairs}\(j^{(s)} \gets \text{Extract-Tuple-Pairs}(j, D)\) 
    \State \label{alg3line:SampleRows}\(j^{(a)} \gets \text{Sample-Rows}(D, \text{exclude}=j^{(s)})\) 
    \State \label{alg3line:MergeandShuffle}\(T \gets \text{Merge-and-Shuffle}(j^{(s)}, j^{(a)})\) 
    \State \textbf{return} \(T\) 
\end{algorithmic}
\end{algorithm}

\textbf{Output} (\(O\)): The expected output is a structured JSON annotation that encapsulates detailed error attributes extracted from the table. Each output instance, selected from the annotated collection \(\mathcal{J}\), comprises essential metadata—including row and column identifiers, the specific error type, the erroneous value, and, when applicable, its corresponding correct value. Additionally, it incorporates any constraints or relational dependencies that furnish contextual information about the error.

 

\subsection{TableEG Training Procedure}
\label{sec:trainingprocedure}

We fine-tune~\cite{brown2020language} our TableEG model using the LLaMA3.1-8B base model with LoRA-based adaptation. The training procedure is organized as follows:

\subsubsection{Model and Adaptation:} We employ the LLaMA3.1-8B model as the base, and apply LoRA~\cite{hu2022lora} to its projection layers (\emph{i.e.,} \(\{q\_proj, k\_proj, v\_proj, o\_proj\}\)) to efficiently adapt the model for table-related error generation tasks.
    
\subsubsection{Hyper-parameters:} We fine-tune our model for 3 epochs using LoRA with a rank of 16. The training is performed with a per-device batch size of 1 and gradient accumulation over 16 steps to effectively increase the batch size. The optimization process employs the AdamW optimizer with a learning rate of \(1 \times 10^{-4}\) and a weight decay of 0.02.
    
\subsubsection{Training Data and Tasks:} We construct the training set from error-labeled datasets (refer to Section~\ref{sec:datasets}) with a 90:10 split (90\% for training). The training tasks include error generation, error detection, and error correction. Additionally, we incorporate synthetic tasks from Table-GPT~\cite{li2024table} (\emph{e.g.,} column/row augmentation, swapping, filtering, sorting, table summarization, header matching, and natural language to SQL conversion) to enhance the model’s understanding of two-dimensional table structures.
    
\subsubsection{Computational Resources:} The training is performed on a distributed cluster equipped with four NVIDIA RTX A6000 GPUs (each with 48GB of memory) to meet the high computational demands of fine-tuning. The implementation is based on PyTorch 2.4.1 with CUDA 12.2.

These components collectively ensure that the fine-tuned TableEG model can effectively learn and generate realistic errors in tabular data.

\subsection{Error Generator}
\label{sec:errorgenerator}

Before deploying TableEG for error generation on user-supplied clean datasets, two practical challenges must be addressed. First, due to the token limitations of large language models, it is infeasible to feed an entire table into the model; hence, the clean table must be partitioned into representative sub-tables. Second, the model may repeatedly target the same cell, causing overlapping modifications. To mitigate these issues and ensure robust error generation, we present the following algorithm~\ref{alg:tableeg_error_generation}, which integrates prompt construction, sub-table sampling, and error application in a controlled manner.

\begin{algorithm}[htbp]
\caption{Error Generation Procedure using TableEG}
\label{alg:tableeg_error_generation}
\begin{algorithmic}[1]
    \Statex \textbf{Input}: Clean dataset \(D^*\), error ratio \(\rho\), error type distribution ratio \(E_r\), TableEG model $f_{\theta}$
    \Statex \textbf{Output}: Generated dataset \(G\)

    \State \label{alg5:line1} $G=D^*$ 
    \State \label{alg5:line2} \(N \gets r(D^*) \times c(D^*)\) 
    \State \label{alg5:line3} \(N_e \gets \lfloor \rho \cdot N \rfloor\) 
    \State \label{alg5:line4} \(\{N_e(e)\}_{e \in E} \gets \text{Distribute}(N_e, E_r)\)
    \State \label{alg5:line5} Used cell set \(U \gets \emptyset\)
    \While{\(\sum_{e \in E} n_e < N_e\)} \label{alg5:line6}
        \State \label{alg5:line7} \(T_i \gets \text{SelectSubTable}(D^*)\) 
        \State \label{alg5:line8} \(I \gets \text{Generate-Instruction}(EGT, e)\)  \Comment{Algorithm~\ref{alg:generate_instruction}}
        \State \label{alg5:line9} \(\hat{O} \gets f_{\theta}(I,T_i)\) 
        \State \label{alg5:line10} Parse \(\hat{O}\) to extract error cell: row \(r\), column \(c\), error value \(\hat{v}\), and correct value \(v\)
        \If{\((r, c) \notin U\) and \(\hat{v} \neq v\)} \label{alg5:line11} 
            \State \label{alg5:line12} \(U \gets U \cup \{(r, c)\}\)
            \State \label{alg5:line13} \(n_e \gets n_e + 1\)
            \State \label{alg5:line14} \(G \gets \text{ApplyErrors}(G, \hat{O})\) 
        \EndIf
    \EndWhile
    \State \label{alg5:line17} \textbf{return} \(G\)
\end{algorithmic}
\end{algorithm}

In our error generation procedure (Algorithm~\ref{alg:tableeg_error_generation}), we first initialize the generated dirty data $G$ to clean data $D^*$ and compute the total number of cells \(N\) in the clean table \(D^*\) by multiplying its number of rows and columns (Line~\ref{alg5:line1}-~\ref{alg5:line2}). We then determine the target number of errors \(N_e\) by applying the error ratio \(\rho\) to \(N\) (Line~\ref{alg5:line3}). Based on the specified error type distribution ratio \(E_r\), the total error count is allocated across error types (Line~\ref{alg5:line4}). An initially empty set \(U\) is used to record cells that have already been modified (Line~\ref{alg5:line5}) in order to prevent TableEG from repeatedly modifying the same cell and causing overlapping changes. In the subsequent while loop (Line~\ref{alg5:line6}), the algorithm repeatedly samples a random sub-table from \(D^*\) (Line~\ref{alg5:line7}) and generates a \textit{Error Generation Task}(detailed in Section~\ref{task:errorgeneration}) instruction \(I\) that include the error type and task requirements (Line~\ref{alg5:line8}). This instruction is paired with the sampled sub-table to form a prompt that is then fed into the TableEG model for inference, yielding an output \(\hat{O}\) (Line~\ref{alg5:line9}). The output is parsed to extract the target cell coordinates (Line~\ref{alg5:line10}). If the identified cell has not been modified previously, the cell is marked as modified and the error count is updated accordingly (Lines~\ref{alg5:line11}-~\ref{alg5:line13}). The resulting $\hat{O}$ is then used to update the generated dirty dataset $G$ (Line~\ref{alg5:line14}).

\section{Evaluation Strategy}
\label{sec:evaluationstrategy}
To thoroughly assess the effectiveness of our error generation approach, we establish a two-part evaluation strategy that examines both the quality of generated errors and their influence on downstream error detection tasks. 

\subsection{Quality of Generated Errors}
\label{sec:eval_gene}

To assess the quality of the generated errors, we define two categories of evaluation metrics:  


\textbf{Error Pattern Alignment Evaluation}: The generated errors should adhere to realistic transformation patterns observed in real-world datasets. To quantify it, we introduce the Error Pattern Alignment Similarity Score, denoted as $S_{\text{EPA}}$. This metric measures the maximum cosine similarity between the transformation vector of a generated error and those derived from real-world errors. The computation follows Algorithm~\ref{alg:evaluation_generated_error}, where we retrieve the  k -nearest real errors in the embedding space and assess the alignment between the generated and real error transformations. Note that the $\phi_{\theta}(x)$ in Algorithm~\ref{alg:evaluation_generated_error} denotes the hidden layer features extracted from the input text $x$. Specifically, we extract the mean of the hidden states from the $L-3$-th layer where $L$ is the total number of layers in the TableEG model $f_{\theta}$. Here, we choose the hidden layers of TableEG as high-dimensional representations of natural language because these layers undergo multiple rounds of self-attention, allowing them to better capture long-range dependencies within sentences. In contrast, the embedding layer in LLMs maps each input token individually into a low-dimensional continuous vector space, focusing primarily on token-level semantics. Meanwhile, the feed-forward network (FFN) layer merely applies nonlinear transformations to the hidden vectors produced by the Transformer layers, primarily serving as an adaptation mechanism for downstream tasks and thus encoding limited additional information.
\begin{algorithm}[h]
\small
\caption{Computation of $S_{\text{EPA}}$}
\label{alg:evaluation_generated_error}
\begin{algorithmic}[1]
    \Statex \textbf{Input}: Generated error dataset $G$, Real error dataset $D$, Number of nearest neighbors $k$, TableEG model without the last two layers $\phi_{\theta}$.
    \Statex \textbf{Output}: Mean error pattern alignment similarity score ($S_{\text{EPA}}$).
    \State $R \gets \text{Build-Real-Dataset}(D,\phi_{\theta})$
    \State $S_{\text{EPA}} = []$
    \For{each $(G^r_c,G^g_e) \in G$}
        \State $\phi_{\theta}(G^r_c) \gets \text{Text-to-Embedding}(G^r_c)$
        \State $\phi_{\theta}(G^g_e) \gets \text{Text-to-Embedding}(G^g_e)$

        \State $indices \gets \text{Find-KNNs}(G^r_c, R, k)$

        \State $\phi_{\theta}(D^r_c), \ \phi_{\theta}(D^g_e) \gets \{R,indices\}$ 
        
        \State $\delta_{real} = (\phi_{\theta}(D^r_c) - \phi_{\theta}(D^g_e))$

        \State $\delta_{gen} = (\phi_{\theta}(G^r_c) - \phi_{\theta}(G^g_e))$

        \State $similarities = \{ 1 - \text{CosineSimilarity}(\delta_{real}, \delta_{gen})\}$

        \State Append $\max(similarities)$ to $S_{\text{EPA}}$
    \EndFor

    \State \textbf{return} $\text{mean} \ (S_{\text{EPA}})$
\end{algorithmic}
\end{algorithm}

\textbf{Error Distribution Alignment Evaluation}: The generated errors should exhibit distributional patterns similar to those observed in real-world datasets. In tabular data, errors often cluster within specific rows or columns, such as missing values appearing in non-mandatory fields or rule violations affecting certain attributes. To evaluate this, we compare the distribution of errors across different columns using metrics such as Jaccard similarity and Jensen-Shannon divergence, which quantify the alignment between the generated error patterns and real-world errors.

\paragraph{Weighted Jaccard Similarity for Column-Level Errors}
Let $\mathcal{C}$ be the set of all columns, with $\mathrm{p}$ and $\mathrm{q}$ denoting the probability distributions of errors across different columns, where $\mathrm{p}(c)$ and $\mathrm{q}(c)$ represent the proportion of real errors and generated errors that occur in column $c$. We compute a local Jaccard value for each column and then average over all columns:
\begin{equation}
J_{\text{col}}^{w} = \frac{1}{|\mathcal{C}|}
\sum_{c \in \mathcal{C}}
\frac{\min\bigl(p(c),\, q(c)\bigr)}
{\max\bigl(p(c),\, q(c)\bigr) + \epsilon },
\end{equation}
where $\epsilon$ is a small constant (\emph{e.g.,} $1 \times 10^{-10}$) to ensure numerical stability. A higher $J_{\text{col}}^{w}$ indicates a closer match between the error distributions of the real and generated datasets in terms of how errors are spread across different columns.

\paragraph{Jensen-Shannon Divergence for Column-Level Errors}
To further analyze the consistency of error type (as mentioned in Section~\ref{sec:errortypes} distributions between generated and real datasets, we compute Kullback-Leibler (KL) Divergence:

\begin{equation}
D_{KL}(X || Y) = \sum_{i} X(i) \log \frac{X(i)}{Y(i)},
\end{equation}
where $X(i)$ and $Y(i)$ represent the normalized frequency distributions of different error types in the real and generated datasets, respectively. However, KL divergence is asymmetric and undefined when $Y(i) = 0$ for any $i$, which can lead to numerical instability. Moreover, its asymmetry makes it less suitable for evaluating how closely the generated error distribution aligns with the real one. To address these issues, we instead use the Jensen-Shannon (JS) divergence, which provides a symmetric and smoothed extension of KL divergence:

\begin{equation}
D_{JS}(X || Y) = \frac{1}{2} D_{KL}(X || Z) + \frac{1}{2} D_{KL}(Y || Z), \quad Z = \frac{X + Y}{2},
\end{equation}
where $Z$ is the average distribution between $X$ and $Y$. Unlike KL divergence, JS divergence measures the similarity between distributions in a more balanced manner by evaluating them within a continuous and uniform reference space defined by $Z$. This property makes JS divergence a more robust metric for comparing error type distributions in real and generated datasets.

\begin{figure*}[htbp]
\centering
\includegraphics[width=0.95\textwidth]{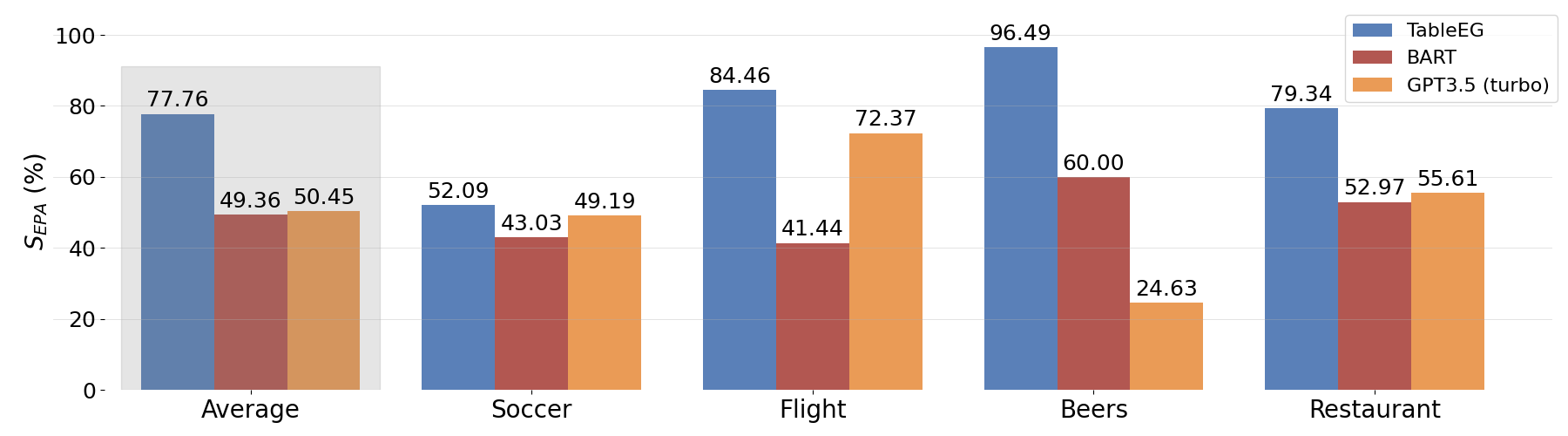}
\caption{Comparison of $S_{EPA}$ (k=20) between TableEG model, BART and GPT3.5 (turbo) across different datasets.}
\label{fig:bar}
\end{figure*}

\subsection{Impact on Error Detection}
\label{sec:error_detection}
Beyond intrinsic quality evaluation, we examine whether the generated errors can be effectively detected by existing error detection algorithms. Specifically, we test multiple error detection models on datasets containing generated errors and compare their detection performance across different error types. This evaluation serves to validate whether our generated errors introduce realistic challenges to error detection methods.  

This evaluation will also explore how well the generated errors align with the inherent characteristics of real-world data and whether they pose realistic challenges for existing detection techniques. We will investigate the detection performance for different error categories (\emph{e.g.,} missing values, rule violations, pattern violations, and outliers) to understand the influence of error types on detection accuracy.

\paragraph{Evaluation Metrics. } 
To evaluate detection performance across different error types, we compute per-category weighted metrics. Given an error type $E_i$, we define its precision $P_{E_i}$, recall $R_{E_i}$, and F1-score $F_{E_i}$ based on the correctly identified errors within this category. The overall weighted scores are then computed as
\begin{align}
P^{w} &= \sum_{i} w_i P_{E_i}, \\
R^{w} &= \sum_{i} w_i R_{E_i}, \\
F^{w} &= \sum_{i} w_i F_{E_i},
\end{align}
where $w_i$ denotes the relative frequency of error type $E_i$ in the dataset. These weighted metrics provide a more comprehensive evaluation by considering the distribution of different error types.



\section{Experiments}
\label{sec:experiments}

In this section, we first introduce the specific configurations of the evaluation experiments (Section~\ref{sec:exp_setup}). We then systematically assess the performance of the TableEG model using the evaluation metrics defined in Section~\ref{sec:evaluationstrategy}. 
Section~\ref{sec:exp_pattern} assesses the semantic and structural similarity between generated and real-world errors, demonstrating that TableEG significantly outperforms BART and GPT-3.5 (Turbo). Section~\ref{sec:exp_distribution} examines the alignment of generated error distributions with real-world data, confirming superior fidelity through weighted Jaccard similarity and Jensen-Shannon divergence. Section~\ref{sec:exp_detection} evaluates the utility of TableEG-generated errors for error detection algorithms, showing strong consistency with real-world benchmarks. These results collectively demonstrate that TableEG can generate errors that closely align with real-world error distributions and structures, making it a robust tool for benchmarking error detection and correction.

All the source code and data are available online.\footnote{\url{https://github.com/viviancircle/TableEG}}

\subsection{Test Configuration}
\label{sec:exp_setup}

We evaluate our trained TableEG model (training details in Section~\ref{sec:trainingprocedure}) by generating errors on four datasets: {Beers}, {Flights}, {Soccer}, and {Restaurant}.  

Among them, {Beers} and {Flights} were seen during training, with 10\% of the data reserved as a test split to assess the model’s learning capability. In addition, we use the {Soccer} and {Restaurant} datasets, which were not exposed to the model during training, to evaluate its generalization performance.

\begin{figure}[t]
\centering
\includegraphics[width=0.47\textwidth]{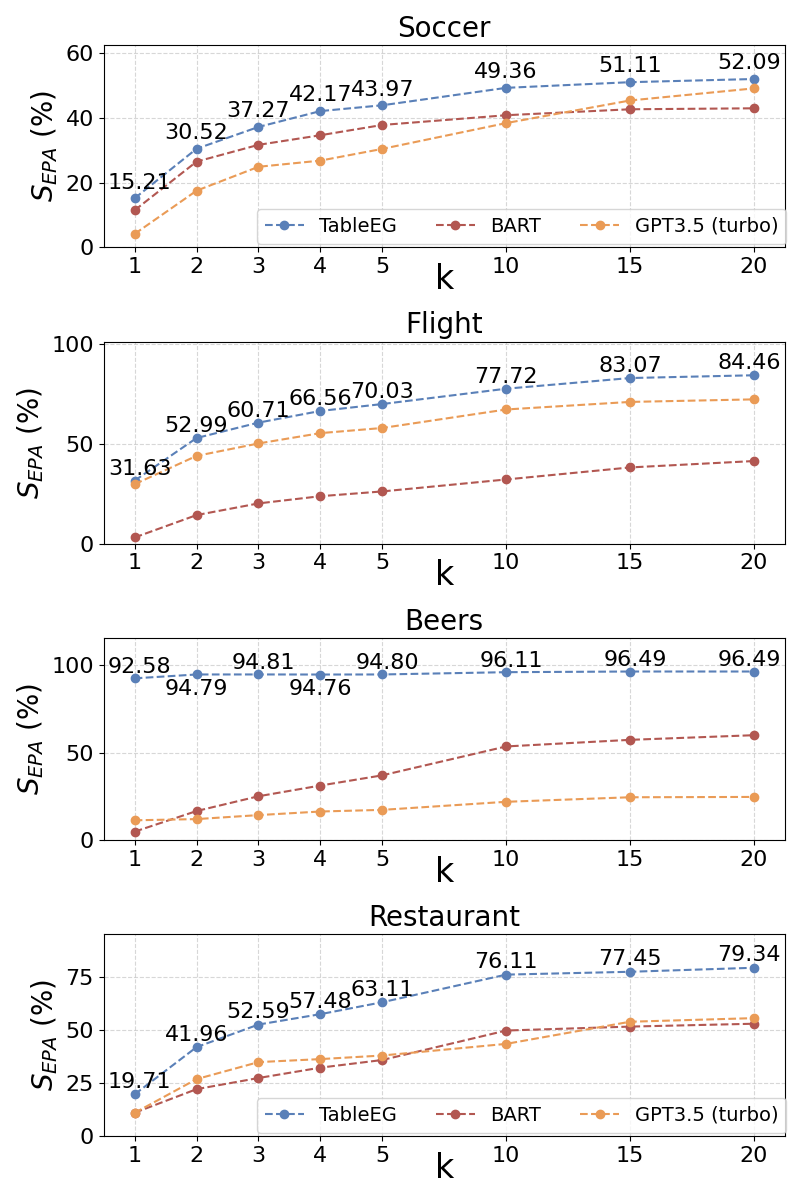}
\caption{Impact of $k$ on $S_{EPA}$ for model TableEG, BART and GPT3.5 (Turbo).}
\label{fig:k}
\end{figure}

\subsection{Error Pattern Alignment Evaluation}

\label{sec:exp_pattern}
To evaluate TableEG’s effectiveness, we compare it against two baselines: the rule-based error generation approach BART and the non-finetuned GPT-3.5 (Turbo). The evaluation metric,  $S_{EPA}$  (\%), measures the similarity between generated errors and real-world data errors. The results are shown in Figures~\ref{fig:bar} and~\ref{fig:k}.

Figure~\ref{fig:bar} summarizes the final $S_{EPA}$  scores at k=20 across all datasets. We can see that TableEG achieves an average $S_{EPA}$ score of 77.76\%, substantially outperforming BART (49.36\%) and GPT-3.5 (50.45\%). Meanwhile, our TableEG demonstrates strong learning ability, achieving high similarity to real errors on the {Beers} dataset, where error patterns are simple and domain-specific. In contrast, GPT-3.5 (Turbo) performs significantly worse than BART, highlighting TableEG’s advantage in specialized domains and the effectiveness of instruction tuning in retaining structural patterns from training data. 
For generalization, while the performance gap between TableEG and the other models narrows on unseen datasets, TableEG still leads. This suggests that TableEG not only learns error patterns effectively in seen data but also transfers them across domains. Its strong learning and stable generalization make TableEG-generated benchmarks more representative of real-world scenarios, enhancing the reliability of error detection and correction evaluations. 

To further analyze the impact of $k$ on error pattern alignment, Figure~\ref{fig:k} illustrates how $S_{EPA}$ varies with different values of $k$ (the number of nearest neighbor vectors considered). 
As $k$ increases, all models exhibit a rise in $S_{EPA}$ scores. This is because a larger $k$ incorporates more nearest-neighbor error patterns, allowing $\delta_{\text{gen}}$ and $\delta_{\text{real}}$ to better approximate the mapping distribution from correct to erroneous values, thereby reducing randomness. 
Furthermore, Regardless of $k$ (from 1 to 20), TableEG consistently outperforms BART and GPT-3.5 (Turbo) across all datasets, demonstrating its stable learning ability and broad applicability. The generated errors closely follow real-world error patterns. 
For a specific dataset, TableEG's advantage over the baselines is smaller on the Soccer dataset. However, as an unseen dataset, Soccer still shows promising overall $S_{EPA}$ growth, suggesting that TableEG effectively generalizes beyond seen data.

\begin{table}[t]
\centering
\caption{Comparison of Weighted Jaccard Similarity ($J_{\text{col}}^w$) and Jensen-Shannon Divergence ($D_{JS}$) across two seen Datasets and three Models, with best-performing values in bold. $\uparrow$ ($\downarrow$)
indicates that higher (lower) values are better.}
\setlength\tabcolsep{11pt}
\renewcommand{\arraystretch}{1.2}
\begin{tabular}{cccc}
\hline
Dataset & Model & $J_{col}^w \ (\uparrow)$ & $D_{JS}\ (\downarrow)$ \\ \hline
\multirow{3}{*}{Flight} & \textbf{TableEG } & \textbf{82.3} & \textbf{7.96} \\
 & BART & 34.69 & 41.96 \\
 & GPT3.5 (Turbo) & 44.68 & 20.43 \\ \hline
\multirow{3}{*}{Beers} & \textbf{TableEG} & \textbf{40.44} & \textbf{28.79} \\
 & BART & 16.04 & 55.85 \\
 & GPT3.5 (Turbo) & 4.96 & 60.81 \\ \hline
\end{tabular}
\label{tab:Jw_DJS}
\end{table}

\begin{table*}[t]
  \centering
  \caption{Performance of different error detection algorithms on real-world dirty data and TableEG-generated dirty data, evaluated using Mean Precision ($P^w$), Recall ($R^w$), and F1-score ($F^w$). $\uparrow$ indicates that higher values are better.}
  \setlength\tabcolsep{6pt}
  \renewcommand{\arraystretch}{1.1}
  \label{tab:detection}
  \begin{threeparttable}     
  \begin{tabular}{lccccccccccccc}
    \toprule
    \multirow{2}{*}{Algorithm} & \multirow{2}{*}{Metrics} & \multicolumn{2}{c}{Flight} & \multicolumn{2}{c}{Movie} & \multicolumn{2}{c}{Company} & \multicolumn{2}{c}{Marketing} \\
    \cmidrule(lr){3-4} \cmidrule(lr){5-6} \cmidrule(lr){7-8} \cmidrule(lr){9-10}
    & & Generated & Real Error & Generated & Real Error & Generated & Real Error & Generated & Real Error \\
    \midrule
    \multirow{3}{*}{Raha} 
    & $P^w \ (\uparrow)$ & 0.92  & 0.90  & 0.86  & 0.89  & 0.99  & 1.00  & 0.74  & 0.43  \\
    & $R^w \ (\uparrow)$ & 0.85  & 0.94  & 0.89  & 0.96  & 0.94  & 0.73  & 0.49  & 0.31  \\
    & $F^w \ (\uparrow)$ & 0.88  & 0.81  & 0.87  & 0.92  & 0.96  & 0.84  & 0.59  & 0.36  \\
    \midrule
    \multirow{3}{*}{Holistic} 
    & $P^w \ (\uparrow)$ & 0.40  & 0.46  & 0.1  & 0.07  & 0.06  & 0.45  & n/a$^\#$  & n/a$^\#$  \\
    & $R^w \ (\uparrow)$ & 0.78  & 0.83  & 0.1 & 0.11  & 0.02  & 0.01  & n/a$^\#$  & n/a$^\#$  \\
    & $F^w \ (\uparrow)$ & 0.53  & 0.59  & 0.1  & 0.09  & 0.03  & 0.01  & n/a$^\#$  & n/a$^\#$  \\
    \midrule
    \multirow{3}{*}{Horizon} 
    & $P^w \ (\uparrow)$ & 0.11  & 0.17  & 0.00  & 0.00  & 0.20  & 0.20  & 0.04  & 0.06  \\
    & $R^w \ (\uparrow)$ & 0.54  & 0.8  & 0.01  & 0.03  & 0.83  & 0.36  & 0.12  & 0.14  \\
    & $F^w \ (\uparrow)$ & 0.18  & 0.3  & 0.00  & 0.00  & 0.31  & 0.25  & 0.06  & 0.08 \\
    \bottomrule
  \end{tabular}
  \begin{tablenotes}
  \item[$^\#$] Holistic did not terminate after one day.
  \end{tablenotes}
  \end{threeparttable}
\end{table*}

\subsection{Error Distribution Alignment Evaluation}
\label{sec:exp_distribution}

Table~\ref{tab:Jw_DJS} presents the comparison of TableEG and the two baselines in terms of Error Distribution Alignment. As mentioned in Sectoion~\ref{sec:eval_gene}, $J_{\text{col}}^w$ measures the column-level similarity between real and generated error distributions, with higher values indicating a closer match. 
On the Flight dataset, it can be seen that TableEG achieves the highest  $J_{\text{col}}^w $ (82.3). This suggests that TableEG effectively captures the real-world error distribution in the Flight dataset, while BART and GPT-3.5 fail to generate errors that match the true column-wise distribution. 
On the Beers dataset, TableEG again surpasses both baselines with a $J_{\text{col}}^w$  score of 40.44, more than 2.5× higher than BART (16.04) and nearly 8× higher than GPT-3.5 (4.96).

$D_{JS}$ quantifies the divergence between real and generated error distributions, where lower values indicate better alignment. 
On the Flight dataset, TableEG achieves a much lower $D_{JS}$ (7.96), demonstrating a strong alignment with real error distributions. On the Beers dataset, TableEG once again achieves the lowest divergence ($D_{JS}$ = 28.79), confirming that its generated errors better match real-world distributions.

In summary, These results strongly indicate that TableEG is not only capable of accurately capturing error type distributions but also excels in learning how errors are structured across table columns. Its consistently lower $D_{JS}$ values further emphasize its superior ability to align generated errors with real-world distributions, making it a more reliable choice for generating realistic benchmark datasets.

\subsection{Error Detection Algorithm Evaluation}
\label{sec:exp_detection}

Following the method described in Section~\ref{sec:error_detection}, we use TableEG to generate errors on four clean datasets and evaluate them using three existing error detection algorithms. The detailed comparison of detection results is presented in Table~\ref{tab:detection}.

According to Table~\ref{tab:detection}, we observe that the performance metrics on TableEG-generated errors closely align with those on real-world errors across nearly all datasets and detection algorithms. For Raha, the most effective detection method, precision, recall, and F1-score remain highly consistent between generated and real errors. 
Furthermore, we find that for error detection algorithms like Raha, which demonstrate strong performance and high adaptability to the given datasets, the similarity in performance between real and generated errors is even more pronounced. This further indicates that for detection algorithms well-suited to the dataset, TableEG-generated errors introduce minimal disruption to detection performance.

These findings validate the effectiveness of TableEG in producing synthetic errors that preserve the complexity and structure of real-world data corruption, establishing it as a reliable tool for training and benchmarking error detection models.
\section{Related Work}
\label{sec:relatedwork}

Ensuring data quality in relational databases and tabular data processing remains a significant challenge, as errors like missing values, pattern violations, and rule violations can disrupt downstream analytical and machine learning tasks. Traditional rule-based approaches, such as functional dependencies (FDs), uniqueness constraints (UCs), and pattern constraints, have been widely used~\cite{rekatsinas2017holoclean}. However, these methods require extensive manual specification, limiting their adaptability to large-scale, high-dimensional data.
Recent advancements in deep learning and large language models (LLMs) have introduced more flexible solutions for data quality management. Transformer-based models have shown strong generalization across various tasks, including data imputation, error detection, and error correction~\cite{li2024table,deng2022turl}. AI-driven methods further enhance adaptability by learning complex data patterns and reducing manual intervention~\cite{zhu2024relational}. 

We categorize existing research into three key areas: data imputation, error detection, and error repairing, tracing their evolution from rule-based methods to deep learning approaches.


\subsection{Data Imputation}
Data imputation is essential for maintaining data completeness. Traditional methods include statistical techniques (\emph{e.g.,} mean, median, mode~\cite{wei2018missing}), matrix factorization (SVD~\cite{troyanskaya2001missing}, PCA~\cite{oba2003bayesian}), regression-based approaches (LOESS~\cite{cleveland1996smoothing}, IIM~\cite{zhang2019learning}), and nearest-neighbor imputation (kNN~\cite{regression1992introduction}, MIBOS~\cite{wu2012missing}). Constraint-based methods leverage data integrity constraints (MLEM2~\cite{grzymala2005handling}, DERAND~\cite{song2015enriching,song2018enriching}), while multiple imputation techniques (MICE~\cite{van2011mice}, SICE~\cite{khan2020sice}) generate multiple candidate values to improve robustness. However, these approaches often rely on strong assumptions, limiting their adaptability to complex data structures.

Machine learning (ML) and deep learning (DL) have improved imputation by capturing intricate data dependencies. Neural network-based models like RNNI~\cite{rumelhart1986learning} and MLPI~\cite{garcia2010pattern} enhance generalization, while autoencoders (MIDAE~\cite{gondara2018mida}, SDAE~\cite{costa2018missing}), VAEs (HI-VAE~\cite{nazabal2020handling}), and graph neural networks (GRAPE~\cite{you2020handling}, IGRM~\cite{zhong2023data}) provide structural awareness. GAN-based approaches (GAIN~\cite{yoon2018gain}, WGAN~\cite{arjovsky2017wasserstein}) further refine imputation by generating plausible missing values. More recently, large language models (LLMs) like GPT-3~\cite{narayan2022can} and Table-GPT~\cite{li2024table} leverage zero-shot and few-shot learning for imputation, offering strong generalization but facing limitations in handling numerical data. Despite these advancements, DL-based methods remain computationally expensive and less interpretable than traditional approaches.

\subsection{Error Detection}
Error detection is essential for identifying anomalies in relational data. Traditional methods include statistical techniques (\emph{e.g.,} PCA~\cite{shyu2003novel}, dBoost~\cite{pit2016outlier}) and constraint-based approaches enforcing integrity constraints (\emph{e.g.,} FDs~\cite{mandros2017discovering}, CFDs~\cite{fan2008conditional}, DCs~\cite{chu2013discovering}). Additionally, clustering-based methods (\emph{e.g.,} DBSCAN~\cite{ester1996density}, OPTICS~\cite{ankerst1999optics}) and outlier detection techniques (\emph{e.g.,} LOF~\cite{breunig2000lof}, iForest~\cite{liu2008isolation}) have also been widely used to capture anomalies.

Beyond traditional approaches, machine learning (ML)-based approaches enhance detection by integrating multiple detection signals and classifier models (\emph{e.g.,} Raha~\cite{mahdavi2019raha}, HoloDetect~\cite{heidari2019holodetect}). Unsupervised (\emph{e.g.,} Uni-Detect~\cite{wang2019uni}) and active learning techniques (\emph{e.g.,} ED2~\cite{neutatz2019ed2}) further improve adaptability by reducing reliance on labeled data. Deep learning (DL)-based methods advance error detection by employing autoencoders for anomaly detection (\emph{e.g.,} RandNet~\cite{chen2017outlier}, REPEN~\cite{pang2019deep}) and utilizing semi-supervised learning to model normal and erroneous data distributions (\emph{e.g.,} DeepSAD~\cite{ruff2019deep}). Recently, large language models (LLMs) have been fine-tuned for tabular data anomaly detection (\emph{e.g.,} ADTDL~\cite{li2024anomaly}), showing potential despite computational challenges.

\subsection{Error Repairing}
Error repairing aims to restore data integrity by correcting erroneous values. Traditional methods include statistical reasoning (\emph{e.g.,} ERACER~\cite{mayfield2010eracer}, SCARE~\cite{yakout2013don}) and constraint-based approaches enforcing FDs and CFDs (\emph{e.g.,} NADEEF~\cite{ebaid2013nadeef}, Holistic~\cite{chu2013holistic}). Clustering-based strategies leverage data similarity for correction (\emph{e.g.,} DORC~\cite{song2015turn}, DISC~\cite{song2021saving}). 

Recent ML-based methods integrate probabilistic reasoning and external data sources (\emph{e.g.,} HoloClean~\cite{rekatsinas2017holoclean}, BoostClean~\cite{krishnan2017boostclean}) to refine repairs. DL-based approaches enhance repair by learning data dependencies (\emph{e.g.,} DeepClean~\cite{zhang2018deepclean}) and leveraging large-scale contextual embeddings (\emph{e.g.,} LLMClean~\cite{biester2024llmclean}). While these methods improve repair accuracy, they remain computationally expensive, requiring further optimization~\cite{tong2016online}.

\section{Conclusion}
\label{sec:conclusion}

In this paper, we presented TableEG, a framework that leverages instruction-tuned large language models to generate authentic synthetic errors in tabular data. Our approach, built on a structured triplet representation \((I, T, O)\), effectively models the intricate two-dimensional dependencies of tables while producing errors that closely mimic those observed in real-world scenarios. Extensive evaluations on both seen and unseen datasets demonstrate that TableEG not only achieves superior error pattern and distribution alignment compared to existing methods, but also serves as a reliable benchmark for downstream error detection and correction tasks. 
Future work will focus on {adaptive learning strategies} to reduce reliance on user-specified constraints during inference. By leveraging domain characteristics and attribute properties (\emph{e.g.,} numerical vs. textual attributes), the model will autonomously generate high-quality dirty data that closely aligns with real-world errors.





\clearpage

\bibliographystyle{ACM-Reference-Format}
\bibliography{sample}

\end{document}